\documentclass[twocolumn,preprintnumbers,amsmath,amssymb,prl]{revtex4}
\usepackage{graphicx}

\begin{document}
\title{Small parameter for lattice models with strong interaction}
\author{A.N. Rubtsov}
\email{alex@shg.ru} \affiliation{Department of Physics, Moscow
State University, 119992 Moscow, Russia}
\begin{abstract}
Diagram series expansion for lattice models with a localized
nonlinearity can be renormalized so that diagram vertexes become
irreducible vertex parts of certain impurity model. Thus
renormalized series converges well in the very opposite cases of
tight and weak binding and pretends to describe in a regular way
strong-correlated systems with localized interaction. Benchmark
results for the classical $O(N)$ models on a cubic lattice are
presented.
\end{abstract}

\maketitle

One of the key problems of the modern condensed-matter theoretical
physics is the quest for a regular description of the strong
correlations \cite{Fulde}. Roughly speaking, correlations are
strong if the nonlinearity and coupling are comparable, and
therefore perturbative approaches make no sense.

Consider a lattice model with local coupling between cells with a
nonlinearity localized inside each cell. In the two limit cases of
the weak and tight binding a series expansion can be performed in
powers of the respective small parameters \cite{Balesku}. One of a
few regular ways to handle the situation of crossover is to employ
the concept of {\it localized} correlations. It basically means
that interaction of each cell with the rest of the system can
approximately be described as an interaction with certain
non-correlated effective bath \cite{Cowley}. The parameters of
this bath are to be found self-consistently. If there is, say, one
particle per cell, it means that real multi-site problem is
approximated by much simpler single-site one ("impurity problem").
An example of such an approach is the mean-field description of
the classical statistical ensembles of interacting particles. In
this scheme, interaction of a particle with its fluctuating
surrounding is replaced with interaction of a particle with a
constant average field. There is a self-consistent equation for
this field. An accuracy can be dramatically increased for change
the mean-field by Gaussian-bath approximation. In this case, the
surrounding is approximated by certain Gaussian ensemble. It is
important that almost the same argumentation can be presented for
very wide class of systems. Particularly, so-called dynamical
mean-field theory (DMFT) for strongly interacting fermions is
essentially a Gaussian-bath approximation \cite{DMFT}.

Physically, Gaussian-bath approximation consists in the assumption
that the irreducible part of high momenta is localized inside a
single site. The strong point of this assumption is that it is
valid for several very different limits for systems with localized
nonlinearity. The approximation reproduces leading-order results
of the weak- and tight-binding series expansion simultaneously.
Indeed, if the nonlinearity is small, the perturbation theory
explicitly says that correlations are localized in the first-order
of weak-binding approximation. On the other hand, if coupling goes
to zero, sites become almost isolated and correlations are
obviously localized, no matter how strong the nonlinearity is.
Once the two opposite limits are described well, their crossover
can be expected to be also somehow depicted. There is some more
justifications for the scheme, particularly it becomes exact for a
system of an infinite dimensionality.

Since an analysis of proper impurity problem gives a good
understanding of the properties of strongly-correlated systems, it
is desirable to construct a perturbation theory starting from the
solution of an impurity problem as a zeroth approximation. A
theory of this kind is presented in this paper for lattice models
with a localized nonlinearity. We exactly renormalize diagram
expansion so that diagram vertexes become irreducible vertex parts
of certain impurity problem. An important peculiarity of the
proposed approach is that a good convergence is achieved both in
tight-binding and weak-binding situations. As an example, we
present the results obtained for quantum Ising model in a
transverse field.


We consider statistics of the real vector field $\phi$ on a
discrete lattice with a localized nonlinearity $\Theta(\phi)$ and
the dispersion law $\Omega$. In general, lattice potential can
depend on a number of the lattice site.
Through the article we number sites by $x$ whereas components of
the on-site field $\phi_x$ are numbered by $j$, statistical
average is denoted by triangle brackets:
\begin{eqnarray}\label{Zphi}
    &<...>=Z^{-1} \int ... ~ e^{-\beta E(\phi)} [D\phi]\\    \nonumber
    &Z=\int e^{-\beta E(\phi)} [D\phi]\\    \nonumber
    &E(\phi)=\sum_x \Theta_x(\phi_x) + \frac{1}{2} (\phi
    \Omega \phi).
\end{eqnarray}
 Scalar
product $(\phi \Omega \phi)$ is a sum over both indices: $\sum_{x
x' j j'} \phi_{xj} \Omega_{x x' j j'} \phi_{x'j'}$; below we also
use external product denoted by the dot: $\phi \cdot \phi$. The
notation $(\phi_x \Omega_{xx} \phi_x)$ is used for a scalar
product in the on-site subspace: $(\phi_x \Omega_{xx}
\phi_x)\equiv \sum_{j j'} \phi_{xj} \Omega_{x x j j'} \phi_{xj'}$.

Consider an auxiliary ensemble with energy
\begin{equation}\label{auxil}
E^{aux}(\phi)=\sum_x \Theta_x(\phi_x)+\frac{1}{2} \left(\phi A
\phi\right)+ (\phi^0 (\Omega-A) \phi).
\end{equation}
We introduced here vector quantity $\phi^{0}$ and a Hermitian
tensor $A$ having non-zero terms only at the same $x$-indices:
$A_{xx'}=0$ for $x\neq x'$ (through the text we call such
quantities '$x$-diagonal'). The values of $A$ and $\phi^0$ are not
specified for a while.

Since $A$ is $x$-diagonal, site oscillators of the auxiliary
ensemble are uncoupled (one can see that $E^{aux}(\phi)$ splits to
a sum of independent single-site energies). Thus, properties of
the auxiliary system at each site can be obtained as a solution of
a single-site impurity problem.

Denote an average over the auxiliary ensemble by overline and
introduce normalized two-point correlator and higher-order
irreducible vertex parts at the sites of auxiliary system:
\begin{widetext}
\begin{eqnarray}\label{gammas}
    &g_{12}=\beta \left(\overline{\phi_1\phi_2}-\overline{\phi_1}~\overline{\phi_2}\right)\\
    \nonumber
    &\gamma^{(3)}_{1'2'3'}=-\beta^3 g_{1'1}^{-1} g_{2'2}^{-1} g_{3'3}^{-1}
    \left(\overline{\phi_1 \phi_2 \phi_3} -
    \overline{\phi_1 \phi_2} ~ \overline{\phi_3} - \overline{\phi_1 \phi_3}~ \overline{\phi_2}
    -\overline{\phi_2 \phi_3} ~ \overline{\phi_1}+2 \overline{\phi_1} ~ \overline{\phi_2} ~
    \overline{\phi_3}\right)
    \\ \nonumber &\gamma^{(n)}_{1'..n'}=(-\beta)^n g_{1'1}^{-1} ... g_{n'n}^{-1} \cdot ({\rm irreducible ~part ~of ~the}~
    n{\rm-th~moment}).
\end{eqnarray}
\end{widetext}
To simplify the notation, we omit site index $x$ and write numbers
as subscripts for the on-site coordinates here.

It should be noted that mean-field and Gaussian-bath
approximations can be understood as approximations of the system
(\ref{Zphi}) by (\ref{auxil}) with a self-consistent requirement
$\phi^0=\overline{\phi}$ and certain choice of $A_{xx}$. Indeed,
to obtain the mean-field scheme, one should, first, replace all
$\phi$ in $E(\phi)$ with its average for all cells except just a
single one, and, second, use an average over this particular cell
as a guess for $<\phi>$. It is easy to check that the  mean-field
corresponds to the simple choice $A_{xx}=\Omega_{xx}$.
Gaussian-bath approximation also reduces the system to a
single-site problem. The difference is that the remaining degrees
of freedom are not frozen in its average positions, but
approximated with harmonic oscillators, forming Gaussian bath.
These Gaussian degrees of freedom can be integrated out, giving
again single-site energy like in (\ref{auxil}) with certain $A$.
An approximation for the single-site system by a harmonic
oscillator can be found from the requirement that the
approximation mimics $\overline{\phi_x}$ and $g_{xx}$. Thus
obtained parameters of harmonic oscillators are to be used as in
the next iteration of the self-consistent loop as parameters of
the oscillators forming the bath. These iterations converge to the
point where $x$-diagonal part of the tensor
$\left(g+(\Omega-A)^{-1}\right)^{-1}$ vanish. This is in fact a
condition for $A$ of the Gaussian-bath approximation.

The aim of this paper is to express averages (\ref{Zphi}) via
$\overline{\phi}$, $g$ and $\gamma^{(n)}$ by certain regular
series. The quantities $A$ and $\phi_0$ will determined by certain
self-consistent conditions.

First, we would like to introduce the "dual" variables $f$. Let us
consider the basis where $g (\Omega-A) g$ is diagonal and label
the states in this basis by $p$. Define the integration $\int [D
f] \equiv \int d f_{p_1} ...\int d f_{p_n} ...$ in such a way that
the path of integration in each integral depends on a sign of
diagonal element $(g (\Omega-A) g)_{pp}$: $f_p$ goes from
$-\infty$ to $\infty$ for $(g (\Omega-A) g)_{pp}<0$ and from $-i
\infty$ to $i \infty$ overwise. With thus defined $\int [D f]$,
the following identity holds:
\begin{widetext}
\begin{equation}\label{Eq3}
    e^{-\frac{1}{2} \beta \left((\phi-\phi^0) (\Omega-A) (\phi-\phi^0)\right)}
    =({\rm factor}) \times \int [Df] e^{-\beta \left(((\phi-\phi^0) g^{-1} f) - \frac{1}{2}
    (f (g (\Omega-A) g)^{-1} f)\right)}.
\end{equation}
In fact, the path of integration over $f_p$ is chosen to deliver
the convergence of integrals here. After this identity is applied,
partition function takes the form
\begin{eqnarray}\label{Zfphi}
    &Z=\int [D\phi] \int [D f] e^{-\beta E(\phi, f)}\\    \nonumber
    &E(\phi,f)=E^{aux}(\phi)
    +((\phi-\phi^0) g^{-1} f)  - \frac{1}{2} (f (g (\Omega-A) g )^{-1} f),
\end{eqnarray}

A set of exact relations between the averages over initial and
dual ensemble can be established by the integration by parts in
(\ref{Zfphi}) with respect to $f$. Particularly,
\begin{eqnarray}\label{ExactRelations}
&g (\Omega-A) <\phi-\phi^0>= <f>;\\   \nonumber &g (\Omega-A)
<(\phi-\phi^0) \cdot (\phi-\phi^0)>(\Omega-A)g = <f \cdot
f>+\beta^{-1} g (\Omega-A) g.
\end{eqnarray}

\end{widetext}

The idea of an introduction of new variables is that energy
(\ref{Zfphi}) does not contain direct coupling between different
sites for the initial variable $\phi$, since $(\phi g^{-1} f)=
\sum_x (\phi_x (g^{-1} f)_x)$. Therefore $\phi_x$ can be
integrated out at each site, yielding an an expression for the
energy in dual variables $E(f)$:
\begin{eqnarray}\label{Zf}
    &E(f)=\sum_x V_x(f_x)+\frac{1}{2} \left(f   \tilde{\Omega}  f\right)\\
    \nonumber
    &\tilde{\Omega}=-g^{-1}-g^{-1}(\Omega-A)^{-1}g^{-1}
\end{eqnarray}
This expression is formally very similar to an initial system
(\ref{Zphi}). Physical difference comes, first, from the
possibility to choose $A$ and $\phi^0$ in certain optimal way,
and, second, from the the non-trivial way of integration over $f$
so that $f$ has an imaginary part. In particular, $<f_{xj}^2>$ can
be negative or equal zero.

Properties of the dual potential $V(f)$ are determined by the
cumulant expansion (\ref{auxil}) for an auxiliary system. The last
term in (\ref{Zf}) is chosen in such a way that the second
derivative of $V$ vanishes; other derivatives are
$V^{(1)}=(\overline{\phi}-\phi^0) g^{-1}$ and $-\beta
V^{(n)}=\gamma^{(n)}$ for $n\ge 3$. To describe effects due to
nonlinearity, we introduce tensor $\Sigma'$ defined by equation
\begin{equation}\label{SigmaG}
    <\phi \cdot \phi>=\beta^{-1} (g^{-1}+\Omega-A+\Sigma')^{-1},
\end{equation}
so that $\Sigma'$ describes  a non-local contribution to the
self-energy of the initial system (\ref{Zphi}). It follows from
the second line of (\ref{ExactRelations}) that dual two-point
correlator can be expressed via $\Sigma$ as follows:
\begin{equation}\label{Sigma}
     <f \cdot f>= g\frac{-\beta^{-1}}{(g^{-1}-\Sigma')^{-1}+(\Omega-A)^{-1}}g.
\end{equation}

All expressions presented above are exact. Now, we construct a
perturbation theory resulting from the series expansion in powers
of $V$. Terms of these series can be expressed by the diagrams
with $\gamma^{(n)}$ standing at vertexes connected with lines
carrying dual two-point correlator $<f \cdot f>$. We consider the
series for $\Sigma'$. It can be found from the expansion of the
left- and right-hand side of (\ref{Sigma}) in powers of $V$ and
$\Sigma'$, respectively. Some of those diagrams are presented in
Figure 1. Each diagram is accompanied by an additional factor
$-\beta^{-1}$ and a numerical coefficient. It can be found from
the above-mentioned expansion of (\ref{Sigma}).

\begin{figure}
\includegraphics[width=\columnwidth]{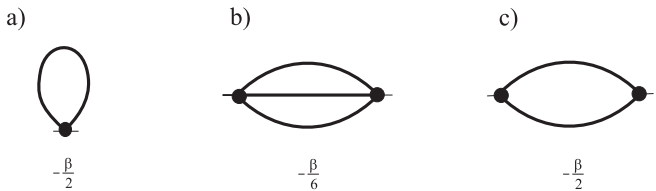}
\caption{Diagrams contributing $\Sigma'$ and corresponding
coefficients. Diagram (a) is small if condition (11) is fulfilled.
Diagrams (b) and (c) describe av $\alpha^2$ correction to
$\Sigma'$; diagram (c) vanishes for an unordered state of $O(N)$
model. }
\end{figure}

Formally, single-lag vertexes can occur in the diagrams due to the
presence of $V^{(1)}$ in the dual energy (\ref{Zf}). We require
such $\phi^0$ that $<f>=0$ to get rid of them.  Then the first
line of (\ref{ExactRelations}) requires
\begin{equation}\label{f0}
    <\phi>=\phi^0.
\end{equation}

Since $\gamma^{(n)}$ of any order can in principle appear in the
series, a formal small parameter should be introduced to group the
terms properly. Consider a special case of a scalar field with
small nonlinearity: $U(\phi)=\alpha \phi^4$, $\alpha \to +0$.
Estimate how $\alpha$ appears in high derivatives of $V(f)$:
$\gamma^{(2 n)}\propto \alpha^n$ and $\gamma^{(2 n-1)}\propto
\alpha^n \phi_0$. We propose to group diagrams with respect to
their formal smallness in powers of $\alpha$. This is clearly a
good choice for a weak-binding limit. Consider the opposite case
of tight-binding $\Omega\to 0$ and suppose that our choice of $A$
delivers also $(A-\Omega)\to 0$ (note that mean-field and
Gaussian-bath schemes satisfy this condition). It is obvious from
(\ref{Zf}) that bare two-point correlator appears to be a small
quantity in this case. Since the number of lines in a diagram is
roughly proportional to its $\alpha$-order, $\alpha$ again appears
to be a good small parameter.  We conclude that {\it series
expansion converges well both in tight-binding and weak-binding
limits} and therefore expect that the theory is suitable for a
crossover situation.

Formally, series expansion can be constructed for any $A$, but an
appropriate choice should result to better convergence. It is
useful to require
\begin{equation}\label{ff0}
   \beta^{-1} g=<(\phi-\phi_0)\cdot(\phi-\phi_0)>_{xx}.
\end{equation}
One can formally obtain Gaussian-bath approximation as a result of
Gaussian approximation for $<f\cdot f>$ in (\ref{Sigma}), combined
with requirements (\ref{f0},\ref{ff0}). Gaussian approximation
means $\alpha^0$ order of the theory. Higher terms of an expansion
in powers of  $\alpha$ improve the result and, in particular,
bring non-local correlations on the scene.

To demonstrate capabilities of the method and to give more
technical details, we present here the results obtained in an
$\alpha^2$-approximation for classical $O(N)$ models at 3D cubic
lattice with the nearest-neighbor coupling. Potential energy is
\begin{equation}\label{Ising}
    -\sum_{<xx'>} (\phi_x \phi_{x'}),
\end{equation}
where $\phi_x$ is an $N$-component vector of the constrained
length $||\phi_x||^2=(\phi_x \phi_x)=N$. In the previous notation,
dispersion law is $\Omega_k=-(\cos k_x+\cos k_y +\cos k_z)$,
whereas the constrain makes the system nonlinear.
 Sum is over all pairs of the
nearest neighbors. Cases of $N=1,2,3$ correspond to Ising, $xy$,
and Heisenberg models, respectively. In three dimensions, system
(\ref{Ising}) shows a second-order phase transition at finite
temperature for $N$. The model is extensively studied. One can
note Monte Carlo simulations \cite{HeisenbergMC, Ising},
high-temperature expansion \cite{HT} (in our terminology, this is
in fact tight-binding approach), renormalisation-group methods
\cite{RG} etc. Here we do not pretend to obtain something new for
the model itself. It is used to check if the method works for a
realistic system.

For simplicity, let us consider an unordered state $<\phi>=0$, so
that only even-lag vertexes appear in diagrams,
$\gamma^{(2n+1)=0}$. It can be easily obtained that
$g_{12}=\beta^{-1}\delta_{12}$ (irrespective on $N$) and
$\gamma^{(4)}_{1234}=\gamma_{N}
(\delta_{12}\delta_{34}+\delta_{13}\delta_{24}+\delta_{14}\delta_{23})$
with $\gamma_{N=1}=-2/3, \gamma_{N=2}=-1/2, \gamma_{N=3}=-2/5$.
Here $\delta$ is a Kronecker delta. Tensor structure of $g$,
$\gamma^{(4)}$, as well as $A$, corresponds to the $O(N)$ symmetry
of the problem.

The self-consistent equation for $a$ follows from (\ref{SigmaG})
and (\ref{ff0}):
\begin{equation}\label{loop}
    g=\frac{1}{(2\pi)^3}\int \frac{d^3 k}{g^{-1}+\Omega_k+\Sigma'-A}
\end{equation}

Technically, this equation can be solved iteratively, similarly to
DMFT-loop technique \cite{DMFT}. Start from certain guess for $A$
and $\Sigma'$. For a general case, one should calculate $g(A)$ and
$\gamma^{(n)}(A)$ at this point, but for a particular model
(\ref{Ising}) it is not necessary as $g$ and $\gamma$ do not
depend on $A$, because of the constrain $||\phi_x||^2=N$.
Calculate $<f \cdot f>$ from (\ref{Sigma}), given $g, A$, and
$\Sigma'$. For thus obtained $<f \cdot f>$ and $\gamma^{(n)}$,
calculate new guess for $\Sigma'$ for a given order in $\alpha$
(see below expressions for $\Sigma'$ up to $\alpha^2$). Finally
calculate the right-hand side of (\ref{loop}), given $g, A$, and
new $\Sigma'$. Denote the obtained value by $\tilde{g}$ and take
the quantity $A+\xi (g^{-1}-\tilde{g}^{-1})$ as new guess for $A$,
where $\xi\lesssim 1$ is a numerical factor. Repeat the loop for
new $A$ and $\Sigma'$. For properly chosen initial guess and value
of $\xi$ iterations converge to a fixed point satisfying equation
(\ref{Ising}).

Gaussian-bath approximation corresponds to an assumption
$\Sigma'=0$. Corrections are given by the diagrams presented in
Figure 1. Lines in these diagrams are thick, {i.e.} they
correspond to the renormalized correlator $<f \cdot f>$.
First-order correction would be given by a simple loop shown as
diagram (a) in Figure 1. This diagram corresponds to formula
$-\frac{1}{2\beta} \gamma^{(4)} <f_x \cdot f_x>$. It is however
absent, because $<f_x \cdot f_x>$ equals zero in Gaussian-bath
approximation, as it can be obtained from (\ref{ExactRelations})
and (\ref{loop}). So the correction starts actually from
$\alpha^2$ terms.

Strictly speaking, once high-order corrections are taken into
account, $<f_x \cdot f_x>$ does not vanish anymore. But it remains
small. It can be shown that simple loops does not appear in
diagrams up to $\alpha^3$ order. Consequently, second-order
correction to $\Sigma'$ can be described by just a single diagram
(b) drawn in Figure 1, so that $\Sigma'_{xx'}= -\frac{1}{6 \beta}
\gamma^2 <f_{x} \cdot f_{x'}>^3$. After the sum over internal
indices is taken, it gives $\Sigma'_{xx'}= -\beta^{-1} \gamma^2
(1+N/2) f_{x, x'}^3$, where $f_{x,x'}$ is a diagonal element of
the dual correlator.

\begin{figure*}
\includegraphics[width=16cm]{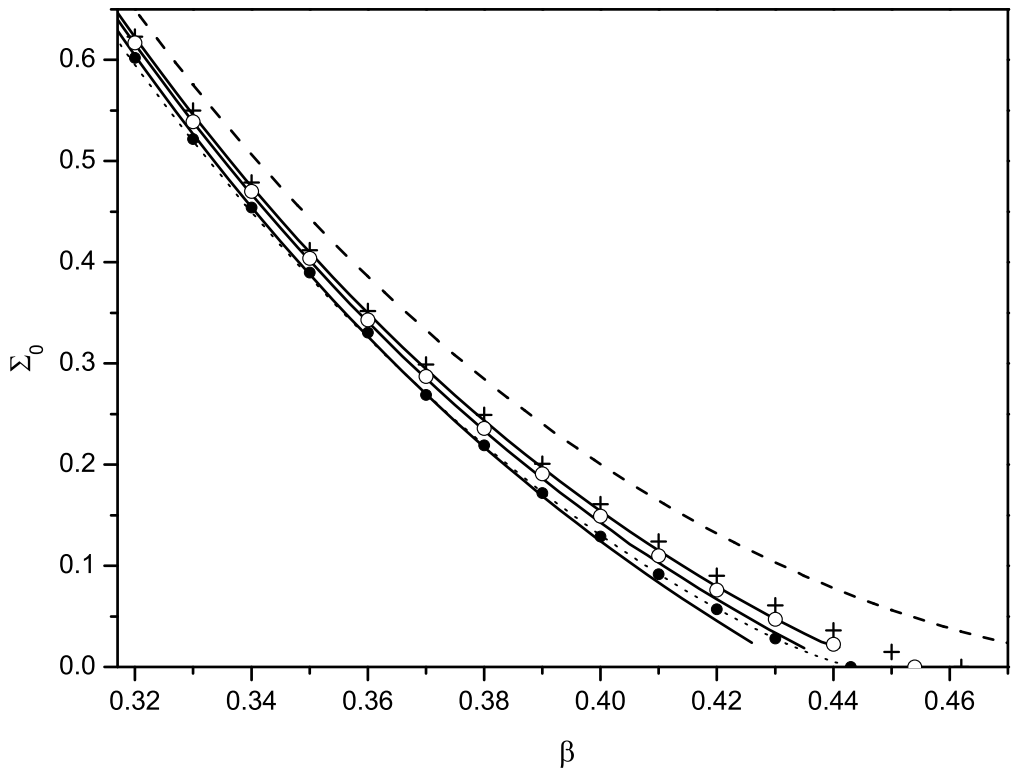}
\caption{ Effective soft-mode dispersion for $O(N)$ model at
$N=1,2,3$. Dashed line: Gaussian-bath approximation. Solid lines:
results for the $\alpha^2$-scheme, from bottom to top: $N=1$
(Ising), $N=2$ (xy), and $N=3$ (Heisenberg). In a narrow region
near critical point the iteration procedure did not converge, and
the lines are not drawn. Symbols are numerical data for $N=1,2,3$:
respectively solid circles, open circles and crosses. Dotted line
is a fit corresponding to a `purely critical' behavior of $<\phi
\cdot \phi>$ in Ising model.}
\end{figure*}

Figure 2 shows data for the effective soft-mode dispersion
$\Sigma_0\equiv(\beta <\phi_{k=0}\cdot\phi_{k=0}>)^{-1}$. Dots are
numerical data. Data for the lines are obtained by formula
$\Sigma_0=g^{-1}+\Omega_{k=0}+\Sigma'-A$. Iteration procedure
converges well everywhere except very narrow regions near critical
point. Since we were not interested in a description of the
critical point itself, we did not put special attention to the
source of this divergence.

 Gaussian-bath approximation predicts the
same dependence for any $N$, as one can observe from (\ref{loop})
with $\Sigma'=0$. This is a serious qualitative drawback, although
the quantitative accuracy is not so bad. So, the dependence of
$\Sigma_0$ on $N$ in the model is related with a non-local part of
correlations, and should be described by higher terms of
$\alpha$-corrections. Indeed, $\alpha^2$-correction describes the
dependence $\Sigma_0(N)$ well. There is also a dramatic increase
of the accuracy in the $\alpha^2$ curves.

For the broken-symmetry state, the theory can be constructed in a
very similar manner. Iteration loop now includes also a
modification of $\phi^0$ to fulfill the condition (\ref{f0}) at a
fixed point. Diagram (c) from Figure 1 appears due to the presence
of $\gamma^{(3)}$. We do not show numerical results here, because
they are very similar to what we obtain for the unordered state.

The entire plot range of Figure 2 lies within a critical region.
For example, all points for Ising model obey critical scaling law
$<\phi_{k=0}^2>\propto (\beta-\beta_c)^{1.24}$ with a good
accuracy, as the dashed line shows. Thus the theory performs well
deep inside the critical region. However, the very vicinity of the
critical point is not correctly described, because we work with
perturbation theory of a finite order. It would be worth to
construct a renormalization-group theory based on the presented
perturbation approach. An attempt to construct a primitive
approach of this kind starting from mean-field theory was
performed several years ago \cite{dualRG}.

There is a comment about an applicability of other approximation
schemes. The system is strongly nonlinear, so weak-binding
approximation is clearly not valid here. On the other hand,
tight-binding description of a low order fails near the phase
transition point because of the critical increase of the
correlation length. For example, second order of the tight-binding
series gives $\Sigma_0=\beta^{-1}(1+4 \beta)^{-1}$. This is very
inaccurate; corresponding curve simply lies out of the plot range
of Figure 2. More sophisticated schemes, for example, cumulant
expansion, also require much higher order to achieve an accuracy
comparable with the presented $\alpha^2$ results. It is also
important to recall that Gaussian-bath approximation becomes exact
in the limit $N\to \infty$, and $1/N$ expansion can be constructed
at this point \cite{1/N}. But $1/N$ approach is essentially based
on a very particular $O(N)$ symmetry of system (\ref{Ising}).
Contrary, our method is developed for a general case and does not
require any special symmetry.

In conclusion, we presented an exact renormalization of the
diagram-series expansion in terms of the self-consistent impurity
model. Renormalized vertexes are irreducible vertex parts of the
impurity model. There is an explicit small parameter in the theory
for weak-binding and tight-binding limit. A worked example of an
$\alpha^2$-approximation for simple 3D models demonstrates that
the method performs well for a strong-correlated situation. The
example models were chosen because of the physical simplicity; the
method itself looks rather general. It would be particularly
important to extend it to the systems of identical quantum
particles. Another challenge is a construction of
renormalization-group theory based on the method.

The work was supported by Dynasty foundation and NWO grant
047.016.005. Author is grateful to M.I. Katsnelson and O.I.Loiko
for their interest to this work.

\end{document}